\documentclass[conference]{IEEEtran}

\usepackage{cite}

\ifCLASSINFOpdf
\else
\fi

\makeatletter
\let\old@ps@headings\ps@headings
\let\old@ps@IEEEtitlepagestyle\ps@IEEEtitlepagestyle
\def\confheader#1{%
  \def\ps@headings{%
    \old@ps@headings%
    \def\@oddhead{\strut\hfill#1\hfill\strut}%
    \def\@evenhead{\strut\hfill#1\hfill\strut}%
  }%
  \def\ps@IEEEtitlepagestyle{%
    \old@ps@IEEEtitlepagestyle%
    \def\@oddhead{\strut\hfill#1\hfill\strut}%
    \def\@evenhead{\strut\hfill#1\hfill\strut}%
  }%
  \ps@headings%
}
\makeatother

\confheader{%
 2016 International Conference on Indoor Positioning and Indoor Navigation (IPIN), 4-7 October 2016, Alcal\a'a de Henares, Spain
}

\usepackage{algorithmicx}
\usepackage{algorithm}
\usepackage{algcompatible}
\usepackage{algpseudocode}
\usepackage{balance}  
\usepackage{graphics} 
\usepackage{times}    
\usepackage{url}      
\usepackage{graphicx,amsmath,mathtools,epstopdf,subfig,caption}
\usepackage{flushend}
\usepackage{fancyhdr,graphicx,lastpage}
\usepackage{placeins}

\def \sys {\textit{CONE}}
\begin{document}
\IEEEoverridecommandlockouts
\IEEEpubid{\makebox[\columnwidth]{978-1-5090-2425-4/16/\$31.00 \copyright 2016 IEEE \hfill} \hspace{\columnsep}\makebox[\columnwidth]{ }}
\title{CONE: Zero-Calibration Accurate Confidence Estimation for Indoor Localization Systems}

\author{\IEEEauthorblockN{Rizanne Elbakly}
\IEEEauthorblockA{Wireless Research Center\\E-JUST, Egypt\\
Email: rizanne.elbakly@ejust.edu.eg}
\and
\IEEEauthorblockN{Moustafa Youssef}
\IEEEauthorblockA{Wireless Research Center\\E-JUST, Egypt\\
Email: moustafa.youssef@ejust.edu.eg}
}

\maketitle

\begin{abstract}
Accurate estimation of the confidence of an indoor localization system is crucial for a number of applications including crowd-sensing applications, map-matching services, and probabilistic location fusion techniques; all of which lead to an enhanced user experience. Current approaches for quantifying the output accuracy of a localization system in real-time either do not provide a distance metric, require an extensive training process, and/or are tailored to a specific localization system.

In this paper, we present the design, implementation, and evaluation of \sys{}: a novel calibration-free accurate confidence estimation system that can work in real-time with \emph{any} location determination system. \sys{} builds on a sound theoretical model that allows it to trade the required user confidence with tight bound on the estimated confidence radius. We also introduce a new metric for evaluating confidence estimation systems that can capture new aspects of their performance.

Evaluation of \sys{} on Android phones in a typical testbed using
the iBeacons BLE technology with a side-by-side comparison
with traditional confidence estimation techniques shows that \sys{} can achieve a consistent median absolute error
difference accuracy of less than 2.7m while estimating the user position more than 80\% of the time within the confidence circle. This is significantly better than the state-of-the-art confidence estimation systems that are tailored to the specific localization system in use. Moreover, \sys{} does not require any calibration and therefore provides a scalable and ubiquitous confidence estimation system for pervasive applications.

\end{abstract}

\IEEEpeerreviewmaketitle

\section{Introduction}
\begin{figure*}[!t]
\centering
    \subfloat[Accuracy Underestimation: When the estimated confidence is larger than the actual localization error.]{
      \includegraphics[width=0.5\linewidth]{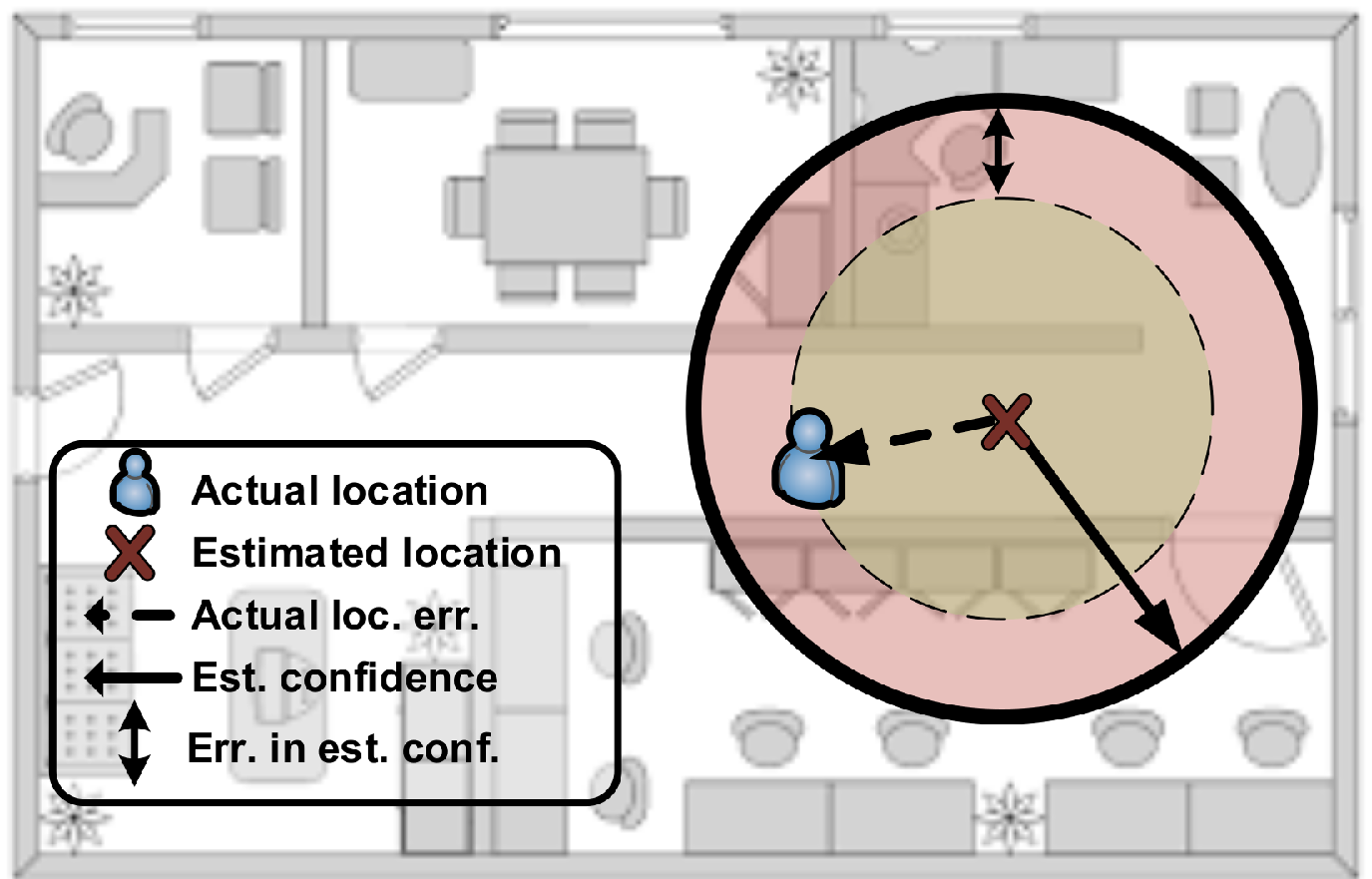}
      \label{fig:error_term_under}
    }
    \subfloat[Accuracy Overestimation: When the estimated confidence is less than the actual localization error.]{
      \includegraphics[width=0.5\linewidth]{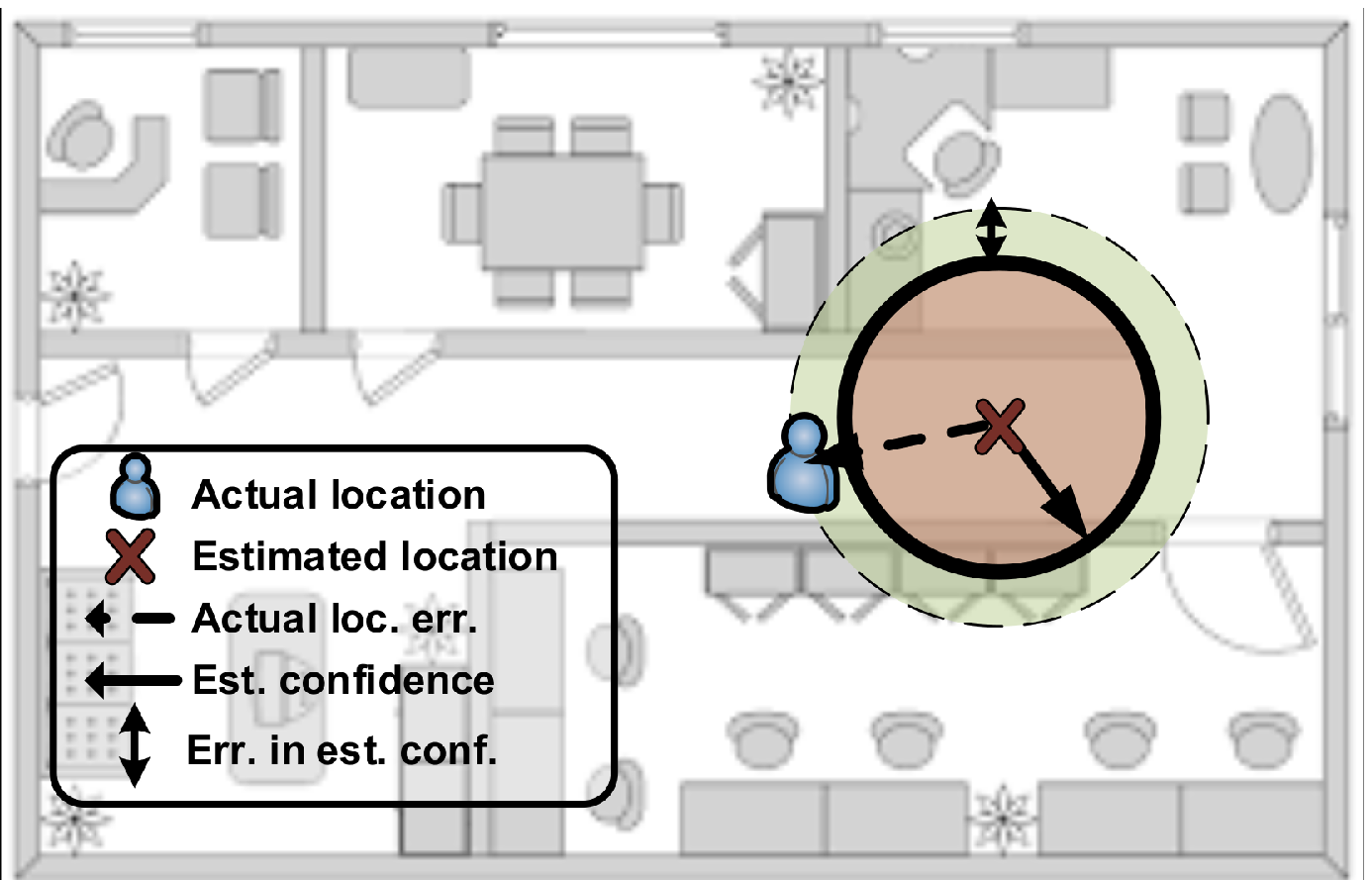}
  \label{fig:error_term_over}
    }
  \caption{Confidence estimation terminology. The solid circle represents the estimated confidence while the dotted circle represents the error in the confidence estimation technique. The figure also shows two important aspects for evaluating a confidence estimation technique: underestimation and overestimation of the actual system accuracy.}
  \label{fig:error_term}
\end{figure*}
The recent advances in wireless technologies, automatic indoor floorplan construction \cite{CrowdInside,elhamshary2015semsense,elhamshary2016transitlabel,elhamshary2014checkinside}, and the wide-spread use of sensor-rich smart phones have sparked research in different indoor localization systems \cite{wang2012no,abdelnassersemanticslam,Horus,alzantot2012uptime,youssef2015towards,Radar,aly2014analysis,sabek2015ace,saeed2014ichnaea,seifeldin2013nuzzer,seifeldin2010deterministic,kosba2009analysis,kosba2012robust,sabek2012multi,abdel2013monophy,youssef2005multivariate,el2010propagation,youssef2006location,eleryan2011synthetic}. Such systems can achieve meter-level accuracy with minimal calibration overhead.
Usually, a localization system is experimentally tested \textbf{\emph{offline}} to quantify its performance against \textbf{\emph{ground truth}} data. While this may be useful for validating the \textbf{\emph{typical error}} of the localization system, it does not help end users understand the \textbf{\emph{real-time error}} in their estimated location. Therefore, a \emph{real-time confidence measure} in the predicted locations would enhance the system usability from end users' perspective. This confidence measure is usually represented as a circle of ambiguity whose center is the estimated user location and radius is the confidence of the estimation (Figure~\ref{fig:error_term}). Such confidence estimation is crucial to enhance the performance of a growing number of applications including crowd-sensing applications~\cite{aly2014map++}, map-matching services~\cite{mohmd_snapnet,aly2015semmatch}, and probabilistic location fusion techniques \cite{kalman}. In addition, it helps users make better use of the provided services~\cite{lemelson2009error}.

To address the error estimation problem, a number of techniques have been proposed in literature \cite{drawil2013gps,moghtadaiee2011indoor,niu2014using,dearman2007exploration,lemelson2009error,moghtadaiee2011indoor}. These systems, however, do not provide the estimated confidence as distance, 
provide a static/constant measure of error, require an extensive calibration process to provide the estimated confidence (which does not scale to large areas and whose accuracy degrades with the continuously varying dynamics of the indoor environment) and/or are tailored to specific localization systems. 
In addition, current systems only quantify the accuracy of the estimated confidence based on the difference between the estimated confidence radius and actual location, which does not capture all aspects of a good confidence measure, e.g. over estimation or under estimation of the estimated confidence.

In this paper, we propose a zero-calibration accurate real-time CONfidence Estimation (CONE) technique for indoor localization systems in the dynamically changing indoor environments. Our proposed method builds on a sound theoretical model for error estimation that can achieve tight error estimates in practical deployment scenarios and works with \textit{any} localization system. We derive our analytical model and show how different error bounds can be achieved based on the required confidence in the user location. We also introduce a new metric for evaluating confidence estimation techniques that can better quantify the system accuracy.

Evaluation of the proposed technique on Android phones in a typical testbed using the iBeacons BLE technology with a side-by-side comparison with traditional confidence estimation techniques shows that \sys{} can achieve a median absolute error difference accuracy of 2.7m while localizing the user within the confidence circle more than 80\% of the time. Compared to a state-of-the-art confidence estimation technique that is tailored to the specific localization system in use, \sys{} provides significantly tighter bounds on the estimated confidence without requiring any information from the localization system nor any calibration overhead.

In summary, our contribution in this paper is three-fold:
\begin{enumerate}
\item We present a theoretical model for estimating the confidence in the estimated user location that can provide different estimation granularities based on the required user-defined confidence. The proposed model does not
require any collection of training data and can be added to \textbf{\emph{any}} location estimation system with no modifications.

\item We propose a novel metric for evaluating confidence estimating systems that can provide fine-grained quantification of the estimated confidence.
\item We implement the proposed confidence estimation system on Android phones and evaluate its performance in a typical indoor environment with a side-by-side comparison with a state-of-the-art system.
\end{enumerate}

\begin{figure}[!t]
\centering
\includegraphics[width=3in] {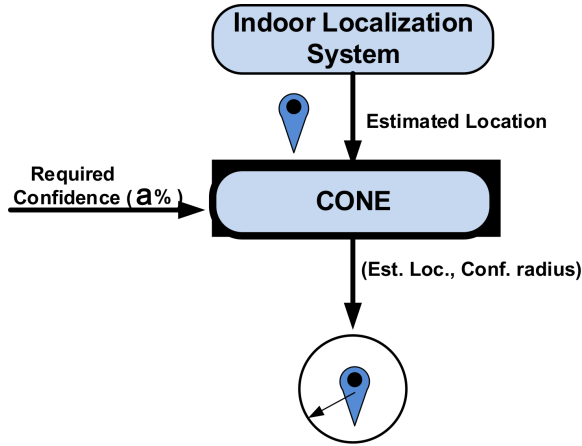}
\caption{Basic system operation: \sys{} combines the estimated location with the required confidence level to estimate the radius of the ambiguity circle, i.e. the circle within which the device is located with the required confidence.
}
\label{fig:cone_oper}
\end{figure}

The rest of the paper is organized as follows: Sections~\ref{sec:acc_est} and \ref{sec:metric} present our proposed confidence estimation method and the new evaluation metric. Section~\ref{sec:eval} evaluates the \sys{} system. Finally, sections \ref{sec:related}  and \ref{sec:conclude} discuss related work and conclude the paper respectively.

\section{Location Confidence Estimation}\label{sec:acc_est}
Figure~\ref{fig:cone_oper} shows the basic system operation. \sys{} can be integrated with any of the current indoor location determination systems without modifications. It combines the estimated location with the required confidence level to estimate the radius of the ambiguity circle, i.e. the circle within which the device is located with the required confidence.

The basic idea \sys{} builds on is simple: if the tracked device is not moving, then the higher the variance \FloatBarrier of the estimated location over time the lower the confidence in its location. In the rest of this section, we provide theoretical analysis to justify this intuition as well as quantify the distance error as a function of the required confidence.

\subsection{Confidence Radius Estimation}

Without loss of generality, we assume that the tracked device is located in 2D.
Let $p_a$ represent the unknown (deterministic) \textbf{actual} user location and $P_e$ be the random variable corresponding to the \textbf{estimated} user location. Let $D_e$ be a random variable that represents the distance error in the estimated location. Therefore, $D_e= \|P_e-p_a\|$. We assume that $D_e$ follows a normal distribution with mean $\mu_e$ and variance $\sigma_e^2$. That is, $D_e \sim \mathcal{N}(\mu_e, \sigma_e)$. This assumption has been verified in our testbed with reasonable accuracy. 

Therefore, given a list of the last estimated $w$ locations ($\mathcal{L}=\{l_1, l_2, ..., l_w\}$), \sys{} estimates the parameters of $D_e$ based on Algorithm~\ref{alg:main}. In particular, the first step is to estimate the center of mass ($\bar{l}$) of the set $\mathcal{L}$. This center of mass should converge to the actual user location as $w \rightarrow \infty$. Then, the distance error ($d_i$) from each location $l_i$ to the to $\bar{l}$ is calculated. Finally, the mean and variance of $D_e$ are estimated as:
\begin{equation}
\mu_e= \frac{1}{w} \sum_{i=1}^{w} d_i
  \label{eq:mean_e}
\end{equation}
\begin{equation}
\sigma_e^2=\frac{1}{w-1} \sum_{i=1}^{w} (d_i-\mu_e)^2
  \label{eq:var_e}
\end{equation}

Once the parameters of $D_e$ are estimated, the required confidence radius ($r$) is estimated based on required confidence level (100$\alpha$\%) using the quantile function:
\begin{equation}
r= \mu_e+\sigma_e \sqrt{2}\, \textrm{erf}^{-1}(2\alpha-1)
  \label{eq:r}
\end{equation}

\begin{figure}[!t]
\centering
\includegraphics[width=3.2in] {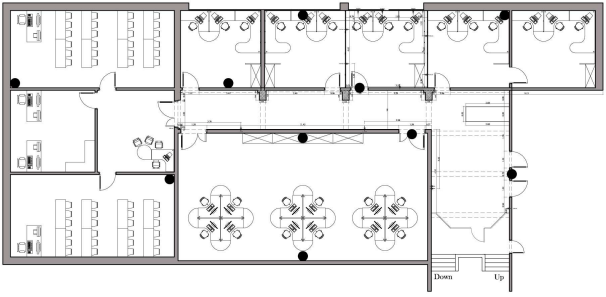}
\caption{Floorplan of the deployment area. The black points represent the beacons locations. The test points were collected uniformly on a grid with a 1m spacing.}
\label{fig:testbed}
\end{figure}

\begin{algorithm}[!t]
\small
\caption{\sys{} Confidence Radius Estimation Algorithm}
\begin{algorithmic}
\State \textbf{Input:} $\mathcal{L}=\{l_1, l_2, ..., l_w\}$ \Comment{Last estimated $w$ locations}
\State \textbf{Input:} $0<\alpha<1$\Comment{Required confidence level}
\State \textbf{Output:} $r$\Comment{Confidence radius}

\State $\bar{l}\gets$ get\_Center\_of\_Mass($\mathcal{L}$)
\For{\textbf{each} $l_i$ \textbf{in} $\mathcal{L}$}
\State $d_i \gets \|l_i-\bar{l}\|$  \Comment{Calculate distance error}
\EndFor
\State $\mu_e \gets \frac{1}{w} \sum_{i=1}^{w} d_i$ \Comment{Mean of $D_e$}
\State $\sigma_e^2 \gets \frac{1}{w-1} \sum_{i=1}^{w} (d_i-\mu_e)^2$ \Comment{Variance of $D_e$}
\State $r \gets \mu_e+\sigma_e \sqrt{2}\, \textrm{erf}^{-1}(2\alpha-1)$ \Comment{Confidence radius}
\end{algorithmic}
\label{alg:main}
\end{algorithm}

\subsection{Discussion}
\sys{} can be integrated with any of the current localization systems as it depends only on the estimated location, without the need for any calibration data.

Since the parameters of the distance error distribution are calculated based on the estimated location, the higher the variance in the estimated location, the higher the variance in the estimated error distance, and hence the higher the estimated confidence radius. This fits our initial intuition.

Similarly, when the user is mobile, the estimated location will be changing frequently, leading to a higher variance and hence to a higher confidence radius. This is intuitive as the user should expect higher ambiguity during movement. Once the person stops moving, her position estimation stabilizes, leading to a reduced confidence radius.

Apart from the required confidence level parameter ($\alpha$), which is specified by the user, \sys{} depends on only one parameter: the window size ($w$) of the last estimated locations. The dependence on only one system parameter makes \sys{} robust to different operation environments. We quantify the effect of $w$ on performance in the evaluation section.

\section{Confidence Evaluation Metrics} \label{sec:metric}
To evaluate the performance of confidence estimation techniques, the \emph{\textbf{absolute}} error difference (AED) has been the only metric usually used in literature~\cite{dearman2007exploration,lemelson2009error,moghtadaiee2011indoor}. The absolute error difference is estimated as the absolute difference between the actual localization error and the estimated error.
Although AED is a good metric, it fails to capture other important performance aspects of confidence estimation techniques. In particular, it is important to quantify how a system overestimates and underestimates the localization system accuracy as depicted in Figure~\ref{fig:error_term}. Note that there is a natural tradeoff between the two aspects: reducing the underestimation error will increase the overestimation error and vice versa. A perfect confidence estimation technique should reduce both to zero, i.e. makes the actual error equal to the estimated error.

To capture these two aspects, we propose to use the \textbf{\emph{signed}} error difference (SED) metric. Specifically, SED is estimated as:
\begin{equation}
\textbf{SED} = \textit{Actual error} - \textit{Estimated error}
\label{eq:SED}
\end{equation}

Hence, a positive SED indicates that the confidence estimation system overestimates the localization system accuracy and that the actual location is outside the predicted confidence circle (Figure~\ref{fig:error_term_over}). On the other hand, a negative SED indicates that the user's actual location is inside the predicted confidence circle, underestimating the system accuracy (Figure~\ref{fig:error_term_under}). 

\section{Performance Evaluation}
\label{sec:eval}

In this section, we evaluate the performance of \sys{} in a typical indoor environment. We start by describing the experimental testbed followed by evaluating the effect of the window size parameter on performance. We finally compare the performance to a typical confidence estimation technique that is tailored to the used localization system.

\subsection{Experimental Testbed}
Our system was evaluated in the second floor of our campus building (Figure~\ref{fig:testbed}). The floor dimensions are 26m by 17m covering corridors, offices, labs, and classrooms. The ceiling height is about 3m.

Our experiments involved the iBeacons technology. Ten Kontact iBeacons were installed throughout the floor as shown in Figure~\ref{fig:testbed} with an average density of one beacon every 44m$^2$. All beacons are installed at the same height of 2.7m.

For evaluating the accuracy of the system, ground truth data was collected using a Samsung Galaxy S5 Android phone. Test points were collected on a uniform grid with a 1m spacing. At each test point, beacons were scanned for 60 seconds in each of the four directions: north, east, south, and west, for a total of four minutes at each point.  

The used localization system is the IncVoronoi calibration-free RF localization system \cite{incvoronoi,elbakly2015calibration} that exploits constraints based on the mutual Received Signal Strength(RSS) relation between each pair of installed RF transmitters to estimate the user's location.
\emph{To reduce the computational overhead, IncVoronoi discretizes the area of interest into a grid and then estimates location based on the the top $k$ candidate grid points.
}
\begin{figure}[!t]
\centering
\includegraphics[width=3.4in] {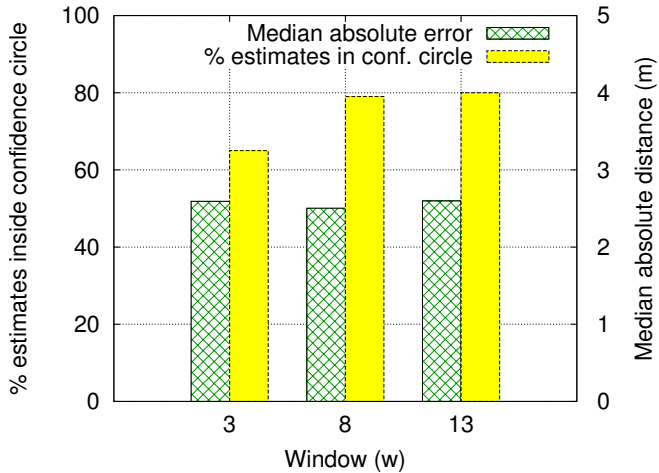}
\caption{Effect of the window size parameter ($w$) on \sys{} performance.}
\label{fig:w}
\end{figure}

\begin{figure*}[!t]
\centering
    \subfloat[\sys{}.\label{fig:var_radius_abs}]{%
    \hspace{-1.7cm}\includegraphics[width=0.5\linewidth]{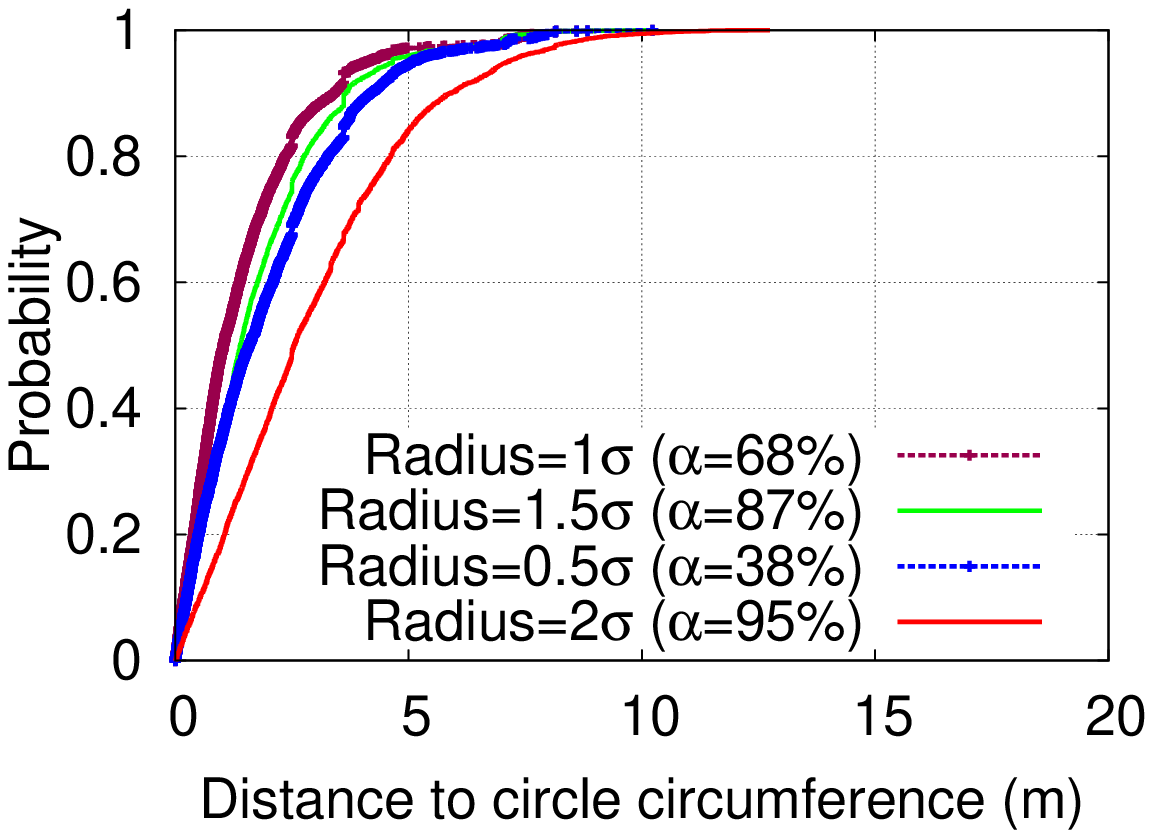}
    }
    \subfloat[GP-Tailored.\label{fig:grid_radius_abs}]{%
    \hspace{-0.65cm}\includegraphics[width=0.5\linewidth]{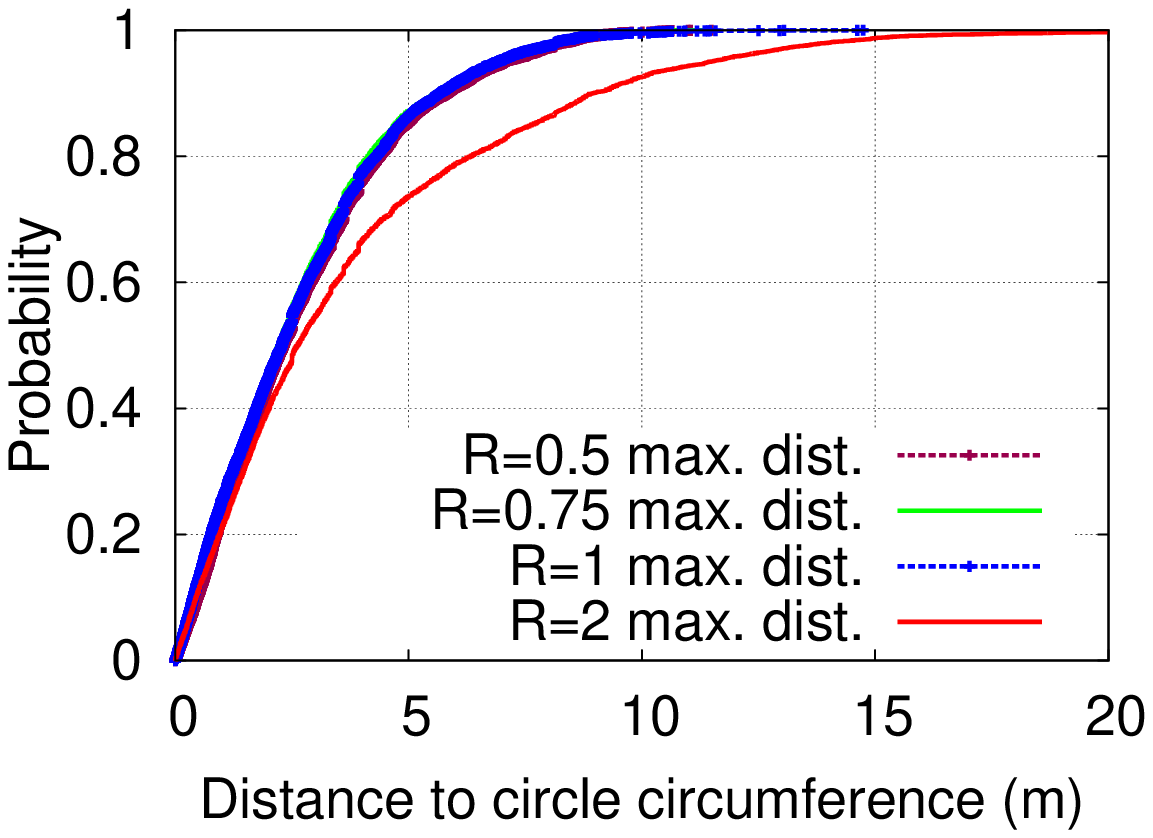}
    }
\caption{Absolute error distance CDF for the different confidence estimation techniques.}
\label{fig:abs}
\end{figure*}

\begin{figure*}[!t]
\centering
    \subfloat[\sys{}.\label{fig:var_radius_sign}]{%
    \hspace{-1.7cm}\includegraphics[width=0.5\linewidth]{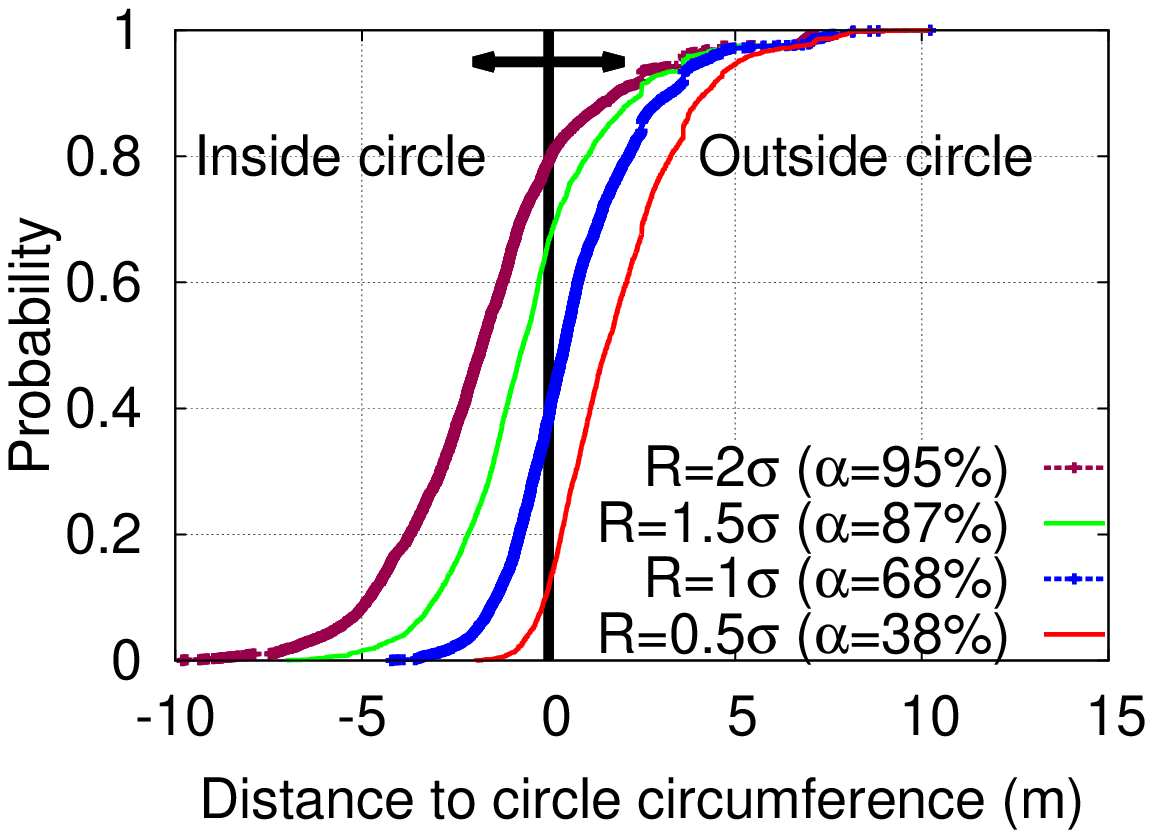}
    }
    \subfloat[GP-Tailored.\label{fig:grid_radius_sign}]{%
    \hspace{-0.65cm}\includegraphics[width=0.5\linewidth]{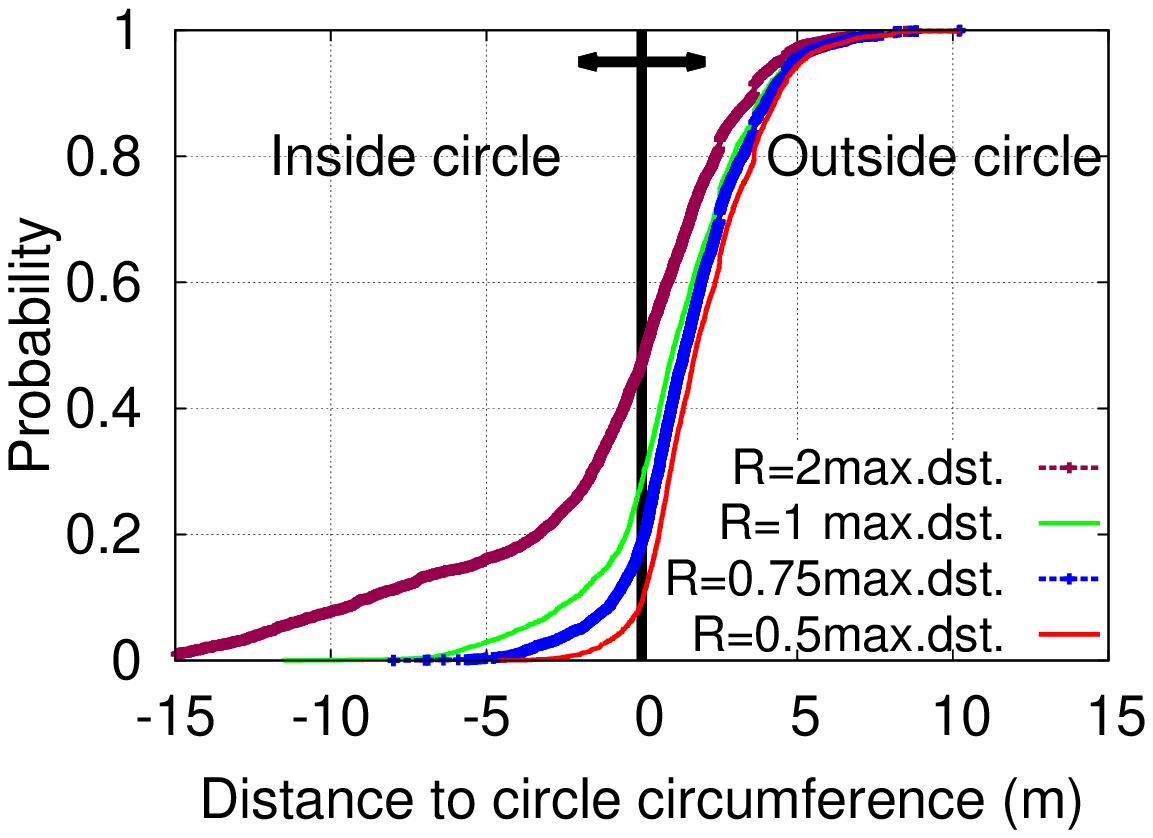}
    }
\caption{Signed error distance CDF for the different confidence estimation techniques. Negative values represent points inside the error circle and vice versa. The best technique should be as close as possible to the line $x=0$.}
\label{fig:signed}
\end{figure*}

\subsection{Effect of the Window Size Parameter on Performance}
Figure~\ref{fig:w} shows the effect of changing the window size of the last estimated locations (parameter $w$) on performance. The figure shows that increasing the window size leads to enhancing the two metrics due to the better estimation of the variance. However, this increases the latency of the estimation process. A value of $w=8$ balances the two factors.
\subsection{Comparison with Other Systems}
In this section, we compare \sys{} to a grid point (GP) based confidence estimation system tailored to the used localization system (we call it ``GP-Tailored'') for different user-specified desired confidence levels. Specifically, since the IncVoronoi localization system \cite{incvoronoi,elbakly2015calibration} discretizes the area of interest into a grid and then estimates location as a function of the top candidate grid points; in the baseline (GP) technique, the radius of the circle is estimated proportionally to the distance between the estimated user location and the furthest grid point within the top $k$ candidate grid points. 

Figures \ref{fig:abs} and ~\ref{fig:signed}  show the absolute and signed distance error CDFs respectively. Table~\ref{tab:confEstComp} summarizes the results. \textbf{\emph{A number of interesting results}} are revealed by the figures: First, for the signed distance error metric, there is a natural tradeoff between underestimating and overestimating the actual system error for the two techniques. The best technique is the closest to the line $x=0$, reflecting the least negative and positive error.

Second, the two metrics reflect different aspects of the performance of the confidence estimation technique: for example, for \sys{}, setting the ambiguity radius to one $\sigma\, (\alpha=68\%)$ gives the best absolute error difference (Figure~\ref{fig:var_radius_abs}) while setting the radius to two $\sigma's\, (\alpha=95\%)$ gives the highest probability of estimated points inside the confidence circle (80\% as in Figure~\ref{fig:var_radius_sign}). Therefore, the designer should take into consideration the different metrics, as opposed to the absolute error metric only -- typically used in literature, when making her decision.

Finally, \sys{} follows the theoretical limit trend with increasing $\sigma$ and provides significantly better results for the two metrics. In particular, \sys{} can estimate the user's actual error with a median absolute error difference less than $2.7$m while estimating the user location within the confidence circle more than 80\% of the time ($\alpha=95\%$). Therefore, it is more rigorous compared to the grid-based method. In addition, unlike the GP-tailored method, it does not require any information related to the internal working of the used localization system. 

\begin{table*}[!t]
\centering
\scalebox{0.9}{
\begin{tabular}{|l|l|l|l|l|} \hline
\textbf{Method}&\textbf{Median abs. error (m)}&\textbf{\% of estimates inside circle}&\textbf{Median +ve dist. (m)}&\textbf{Med. -ve dist. (m)}\\ \hline \hline
\textbf{\sys{} ($R=2\sigma$})&2.54&80\%&1.732&-2.3282\\ \hline
\textbf{GP-Tailored ($R=0.5$ max. dist.)}&2.32&10\%&2.5515&-0.8187\\ \hline
\end{tabular}}
\caption{Comparison between confidence estimation methods. The median +ve and -ve distance error reflect the median error when the estimate is inside or outside the confidence circle respectively.}
\label{tab:confEstComp}
\end{table*}

\section{Related Work}
\label{sec:related}

Recently, researchers have proposed confidence estimation techniques for the different localization systems to improve the usability of their location predictions, e.g. GPS~\cite{drawil2013gps,moghtadaiee2011indoor}, GNSS~\cite{niu2014using} and fingerprinting-based localization techniques~\cite{dearman2007exploration,lemelson2009error,moghtadaiee2011indoor}.

Confidence estimation for the GPS is typically derived from the geometric
dilution of precision, especially the horizontal dilution
of precision (HDOP) \cite{drawil2013gps,moghtadaiee2011indoor,niu2014using}. The HDOP, however does not provide a measure of error in meters but rather a scaling factor based on
the geometry of the visible satellites. An HDOP value of three or less indicates
good satellite geometry and a relatively accurate location estimate \cite{dearman2007exploration}. 
To obtain a confidence measure in meters, in~\cite{drawil2013gps} and~\cite{niu2014using} authors analyze the error characteristics of GPS/GNSS localization systems and build an error model to estimate the localization error.

For indoor localization, confidence estimation techniques typically try to leverage their war-driving/training data to estimate an accuracy measure along with the predicted location. 
For example, in~\cite{dearman2007exploration} authors proposed to maintain a database of locations and their corresponding measured error to estimate the fingerprinting-based systems' accuracy. Such a database is built off-line by running the localization system on a set of measurements with known actual locations and recording the predicted locations and their associated error in the database. Similarly, in~\cite{lemelson2009error}, authors proposed to use the leave-one out method, instead of having separate traces to estimate the error and proposed other techniques to estimate the confidence including fingerprints clustering and signal strength variation.

In comparison to these fingerprinting-based and GPS and GNSS localization systems, \sys{} depends on the estimated location only, without requiring any internal information from the localization systems or extensive calibration, which may change over time affecting the confidence estimation system accuracy. This allows it to work with any location determination system under the dynamically changing indoor environment. 

\section{Conclusion}
\label{sec:conclude}
We presented the design, implementation, and evaluation of \sys{}, an accurate fingerprint-less confidence estimation technique for indoor localization systems. \sys{} works only on the estimated location to quantify the accuracy of the system, and hence can work with any localization system. We derived the accuracy bound as a function of the user desired confidence level as well as proposed a new signed distance error metric that can capture different performance aspects of confidence estimation systems than the traditional absolute distance error metric.

  Evaluation of \sys{} in a typical testbed using the iBeacons technology shows that it can achieve a median distance error of less than 2.7m while estimating the user location within the confidence circle more than 80\% of the time. This is better than a traditional state-of-the-art confidence estimation technique that is tailored to the localization system in use. In addition, \sys{} does not require any internal information about the performance of the localization system nor any calibration overhead.

  Currently, we are expanding \sys{} in different directions including evaluation on other testbeds, integration with other localization systems, deriving a hybrid evaluation metric, among others.

\section{Acknowledgment}
This work is supported in part by NaviRize Inc. and in part by a grant from the Egyptian Information Technology Industry Development Agency (ITIDA).
\bibliographystyle{IEEEtran}

\end{document}